\documentstyle[11pt,pasp,twoside,epsf]{article}
\def\edcomment#1{\iffalse\marginpar{\raggedright\sl#1\/}\else\relax\fi}
\reversemarginpar

\begin{document}
\title{Resolving the X-ray Spectral Paradox: XMM-Newton Spectra of 
Faint Sources in the Lockman Hole}
 \author{Richard E. Griffiths, Andrew Ptak, Takamitsu Miyaji}
\affil{Dept. of Physics, Carnegie Mellon University, 5000 Forbes Ave.,
Pittsburgh, PA 15213-3890, USA}

\begin{abstract}
Using 100ks of XMM-Newton data on the Lockman Hole, 
we show how the X-ray background 'spectral paradox' is 
being resolved. We find the summed spectra of the Type I AGN,
Type II AGN, and the unidentified objects. We conclude that
the hard slope of the XRB is caused predominantly by latter
sources, i.e.  the Type II AGN and unidentified objects.
\end{abstract}

\section{Introduction}

The X-ray background has an effective energy spectral index of 0.4
between 2 and 20 keV, whereas AGNs have steeper spectral indices
(0.8 for Type I). 
One of the major goals of the current generation of X-ray observatories 
has been to identify those source populations which are primarily
responsible for the bulk of the overall energy density, which 
peaks at 30 kev. In particular,
we aim to classify those sources contributing to the source
counts in the range 2 - 10 kev, and to recognize those with the 
hardest spectra. 

\section{Observations and Results}

The XMM-Newton observatory was pointed at the Lockman Hole 
in Apr./May 2000, resulting in a total of $\sim$100 ks. of useful data.

A total of $\sim150$ sources was detected, as reported in Hasinger
et al. 2000 (A\&A, {\bf 365}, L45).
Of these sources, about 50 were previously detected with ROSAT, so that 
about 100 are new detections with XMM-Newton. 

Summed x-ray spectra (Table 1) are shown in Fig. 1 for sources 
with redshifts from optical spectroscopy
(Lehmann et al 2001 A\&A {\bf 371}, 833), and for sources
with no ROSAT counterpart (and hence no optical counterpart as yet).
The summed X-ray spectra of type I AGN, type II AGN, and
objects without optical identification are shown.

\section{Conclusions}

From a deep XMM-Newton observation of the Lockman Hole, and 
using optical identifications established following the 
deep ROSAT survey of the same area, we have found that: 

\vskip -2in

\begin{table}
\vskip -0.3in
\caption{X-ray Spectral Indices}
\begin{tabular}{l|ccccc}
Fit &	N &	Nh	&	Gamma	&		F(2-10)	&	Chisq/dof \\
\hline
  &	 &	$10^{21}$ cm$^{-2}$ &           &                       &               \\
AGN I	& 42  &	0.5 ($<1.1$) &	1.87 (1.80-1.94)  &	2.32e-14 &	632/606 \\
AGN I$^*$	& 27  &	0.4 ($<1.0$) &	1.85 (1.78-1.92)  & 	3.55e-14 &	600/562 \\
AGN II &  5   & 2.6 ($<5.4$) &	1.19 (0.93-1.45)  &	2.68e-15 &	71/71 \\
No ROSAT ID   &	99  & 0.0 ($<1.4$) &	1.49 (1.39-1.68)  &	3.58e-15 &	711/635 \\
No ROSAT ID$^*$ & 40 & 1.8 ($<3.7$) & 1.11 (0.93-1.31)  &	7.06e-15 &	284/309 \\

\end{tabular}
$^* = $ selected to have a $>90\%$ confidence detection in the 5-8 keV
 bandpass.
F(2-10) is the mean flux/galaxy in cgs units.  All AGN II galaxies have significant
detection in the hard band.
N = number of galaxy spectra summed\\
AGN I = sources in Lehmann et al. with class = a-c\\
AGN II = sources in Lehmann et al. with class = d\\

\vskip -1.5in
\end{table}

\begin{figure}
\caption{Spectra Summed by AGN Type\\
{\it top} AGN Type I, {\it middle} AGN Type II, {\it bottom}
unidentified sources, hard-selected}
\plotfiddle{griffiths_richard_fig1.epsi}{4in}{270}{50}{50}{-200pt}{350pt}

\null
\vskip -1.0in
\end{figure}

\vskip 1.3in
(i) Type I AGN have an ave. spectral index of about 1.85
over 2 -- 10 keV

(ii) Type II AGN have an average energy spectral index of 
about 1.2

(iii) sources with no optical spectra (thus far)
have an energy spectral index
of 1.5, or 1.1 for the hard-selected sources (5 - 8 kev)

Extrapolating to higher energies, it is evident that AGNs of 
Type II and  (non-ROSAT) unidentified sources make up the bulk of the 
energy in the XRB.

\acknowledgments

These observations were made using the XMM-Newton X-ray observatory,
built, launched and operated by the European Space Agency with 
NASA participation. These results rely heavily on the analysis
performed and reported by Guenther Hasinger and colleagues
(A\&A 2000 {\bf 365}, L45). 
 
\end{document}